\documentclass[sigconf]{acmart}

\renewcommand\footnotetextcopyrightpermission[1]{}

\usepackage{kotex}
\usepackage{subfigure}
\AtBeginDocument{%
  \providecommand\BibTeX{{%
    \normalfont B\kern-0.5em{\scshape i\kern-0.25em b}\kern-0.8em\TeX}}}

\setcopyright{none}
\copyrightyear{}
\acmYear{}
\acmDOI{}

\acmConference[]{}{}{}
\acmBooktitle{}
\acmPrice{}
\acmISBN{}

\begin{document}

\title{Interpreting Neural Ranking Models using Grad-CAM}


\author{Jaekeol Choi}
\authornotemark{}
\affiliation{%
 \institution{Seoul National University}
  \streetaddress{P.O. Box 1212}
  \city{Seoul}
  \country{Korea}
  \postcode{43017-6221}  
}
\email{jkchoi.naver@navercorp.com}

\author{Jungin Choi}
\affiliation{%
  \institution{Seoul National University}
  \streetaddress{}
  \city{Seoul}
  \country{Korea}}
\email{luvimperfection@gmail.com}

\author{Wonjong Rhee}
\affiliation{%
  \institution{Seoul National University}
  \city{Seoul}
  \country{Korea}
}
\email{wrhee@snu.ac.kr}


\begin{abstract}
Recently, applying deep neural networks in IR has become an important and timely topic. For instance, Neural Ranking Models(NRMs) have shown promising performance compared to the traditional ranking models. However, explaining the ranking results has become even more difficult with NRM due to the complex structure of neural networks. On the other hand, a great deal of research is under progress on Interpretable Machine Learning(IML), including Grad-CAM. Grad-CAM is an attribution method and it can visualize the input regions that contribute to the network's output. In this paper, we adopt Grad-CAM for interpreting the ranking results of NRM. By adopting Grad-CAM, we analyze how each query-document term pair contributes to the matching score for a given pair of query and document. The visualization results provide insights on why a certain document is relevant to the given query. Also, the results show that neural ranking model captures the subtle notion of relevance. Our interpretation method and visualization results can be used for snippet generation and user query-intent analysis.
\end{abstract}



\maketitle

\section{Introduction}

Deep neural network has shown a great promise in various fields including computer vision and natural language processing.
Neural network has a superior predictive power thanks to its complex architecture and large number of parameters, but instead it provides a very low level of explainability. To overcome the limitation of neural network being treated as a "blackbox", research on ML interpretability is actively under progress. While a great deal of research has been done in image tasks, little progress has been made in IR field, especially in ranking tasks.

In this paper, we propose an interpretation method of applying Grad-CAM algorithm to neural ranking model. Grad-CAM \cite{selvaraju2017grad} is a recently proposed interpretation algorithm which measures each input's contribution to the target activation. Since Grad-CAM is applicable to CNN-based architecture, we choose MatchPyramid \cite{pang2016text} as our model of interest. MatchPyramid, one of the interaction-focused model \cite{guo2019deep}, uses query-document interaction matrix as input to the CNN. Each element of the interaction matrix, $M_{ij}$, denotes the similarity between $q_i$, the $i$-th term in query and $d_j$, the $j$-th term in document. Given a pair of query and document, we set the target activation as the final ranking score of the document. By deriving the localization map $L_{Grad-CAM}$, the output of the Grad-CAM algorithm, we measure the contribution of each $(q_i, d_j)$ pair to the ranking score.

We provide explanations for following two questions: 
$i)$"Given a query and a document, which components of the document contribute to the ranking result?"
$ii)$ "Given a query and a set of documents, which characteristics differentiate the ground truth document from the negative documents?" 

For the first question, we extracted {\it effective terms} and {\it filtered terms} from each document, given a query. We conducted an experiment to show that localization map $L$ can be useful to extract the most query-relevant snippet from a document. For the second question, statistical analysis reveals several interesting observations. The ground truth document has higher values for $K$ and $\sum_{i}^{}\sum_{j}^{}L_{ij}$ than the negative documents. $K$ denotes the kurtosis of localization map $L$. $\sum_{i}^{}\sum_{j}^{}L_{ij}$ denotes the sum of contributions of every query-document term pairs within a document.\\

\begin{figure*}[t]
	\centering
	\includegraphics[width=\linewidth]{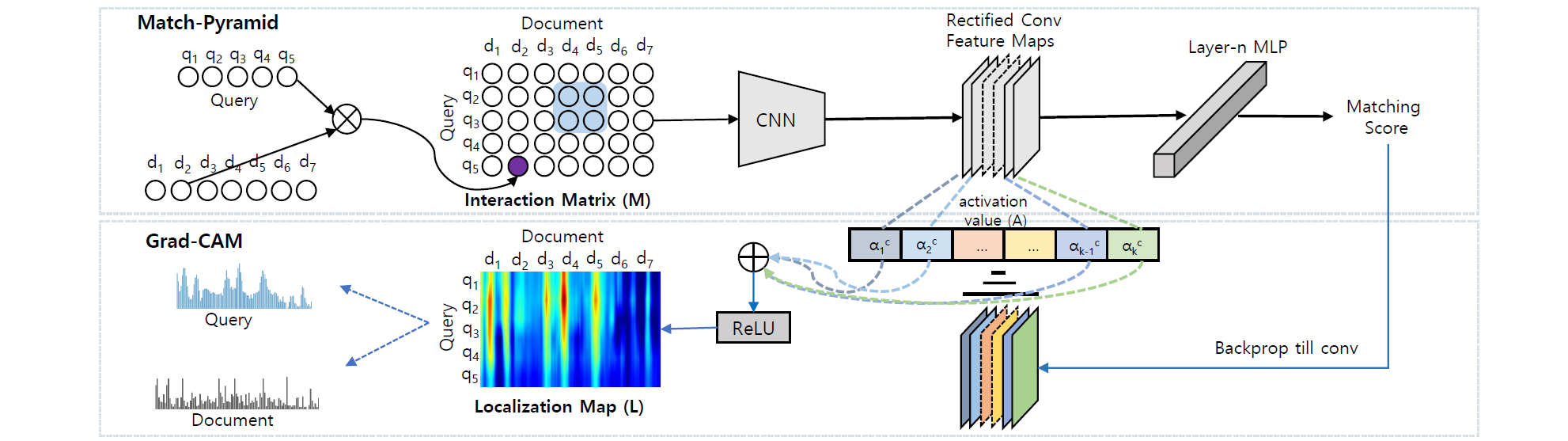}
	\caption{Structure of Match Pyramid model and our interpretation method via Grad-CAM algorithm}
	\label{fig:gradcam}
	\Description{}
\end{figure*}

\section{RELATED WORKS}
Model interpretability method can be categorized into model-agnostic approach and model-introspective approach: Model-agnostic approach \cite{ribeiro2018anchors, singh2018posthoc, verma2019lirme, singh2019exs} tries to approximate the original "blackbox" model by learning an interpretable model. These methods can be applied to any types of model, but are limited to indirect explanations due to approximation. In contrast, model-introspective approach tries to explain the base model's operation by investigating the internal components such as features and gradients. In \cite{fernando2019study}, they investigate the robustness and accuracy of different reference input methods in attempt to utilize DeepSHAP algorithm for NRM interpretability.

We believe that model-introspective approach has its advantage in providing direct explanation of ranking results. For classification tasks, there have been many model-introspective methods \cite{simonyan2013deep, bach2015pixel, selvaraju2017grad} to interpret neural network classifiers. In this respect, we propose a model-introspective NRM interpretation method by adapting Grad-CAM algorithm\cite{selvaraju2017grad}, a verified interpretation technique for image classifier. To the best of our knowledge, our work is the first systematic approach to investigate internal functionality of NRM.

\section{Grad-Cam for Neural Ranking}

Grad-CAM is an attribution method to calculate each input's contribution to the output. Grad-CAM explains the output layer decision by using gradients flowing into the last convolutional layer of the CNN. In classification tasks, Grad-CAM is used to highlight the image regions that highly contribute to the class prediction score. In the context of ranking task, Grad-CAM can be used to explain how each input contributes to the matching score between a query and a document.

Figure \ref{fig:gradcam} shows the entire process of applying Grad-CAM algorithm to MatchPyramid ranking model.
MatchPyramid\cite{pang2016text}, our model of interest, uses a query-document interaction matrix $M$ as the CNN input. Each element of the interaction matrix $M_{ij}$ denotes the similarity between $q_{i}$, the $i$-th term in query and $d_{j}$, the $j$-th term in document. Interaction matrix $M$ enters a CNN, producing high-level matching patterns. The high-level matching patterns are then fed into a multi-layer perceptron to produce the final matching score between a query and a document.

\begin{figure*}
	\subfigure [$L$(top) and $l$(bottom)]{\includegraphics[scale=0.25]{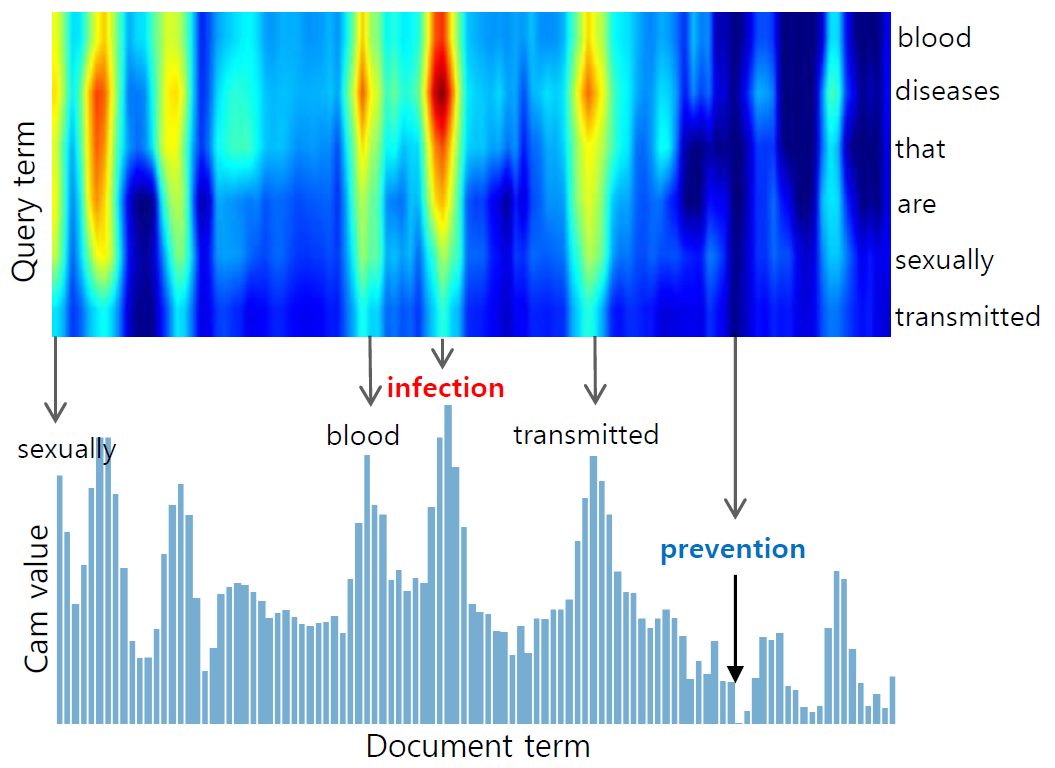}}
	\subfigure [$M$(top) and $m$(bottom)]{\includegraphics[scale=0.25]{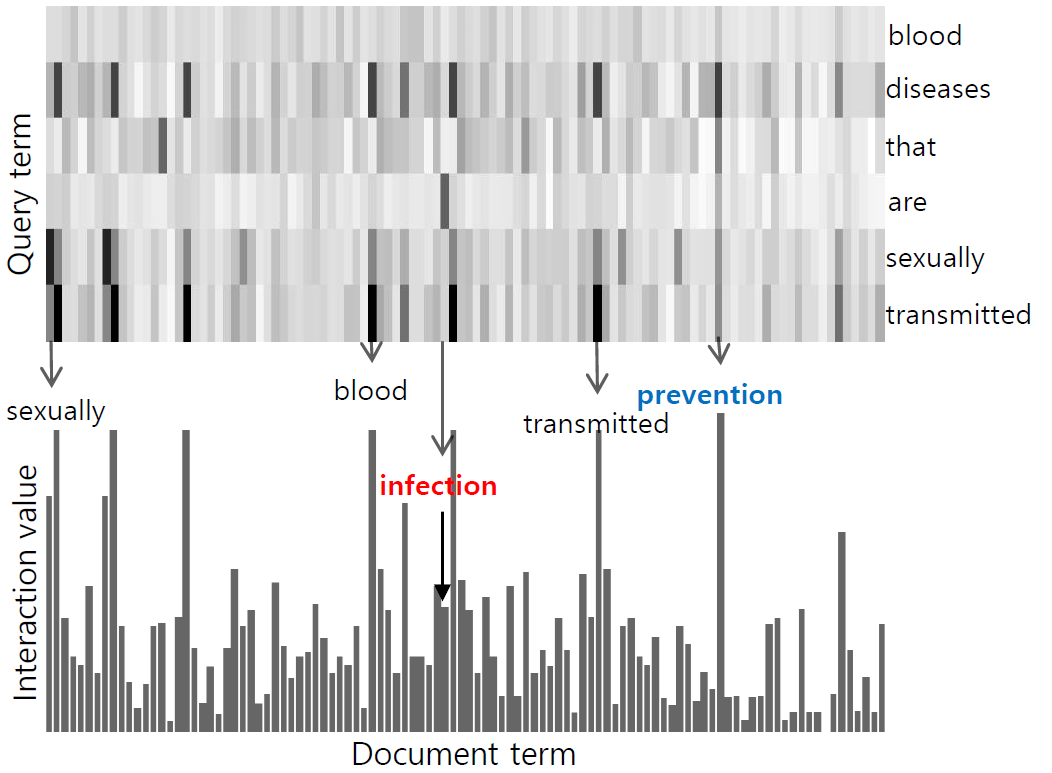}}
	\subfigure [Effective terms(top) and Filtered terms(bottom)]{\includegraphics[scale=0.25]{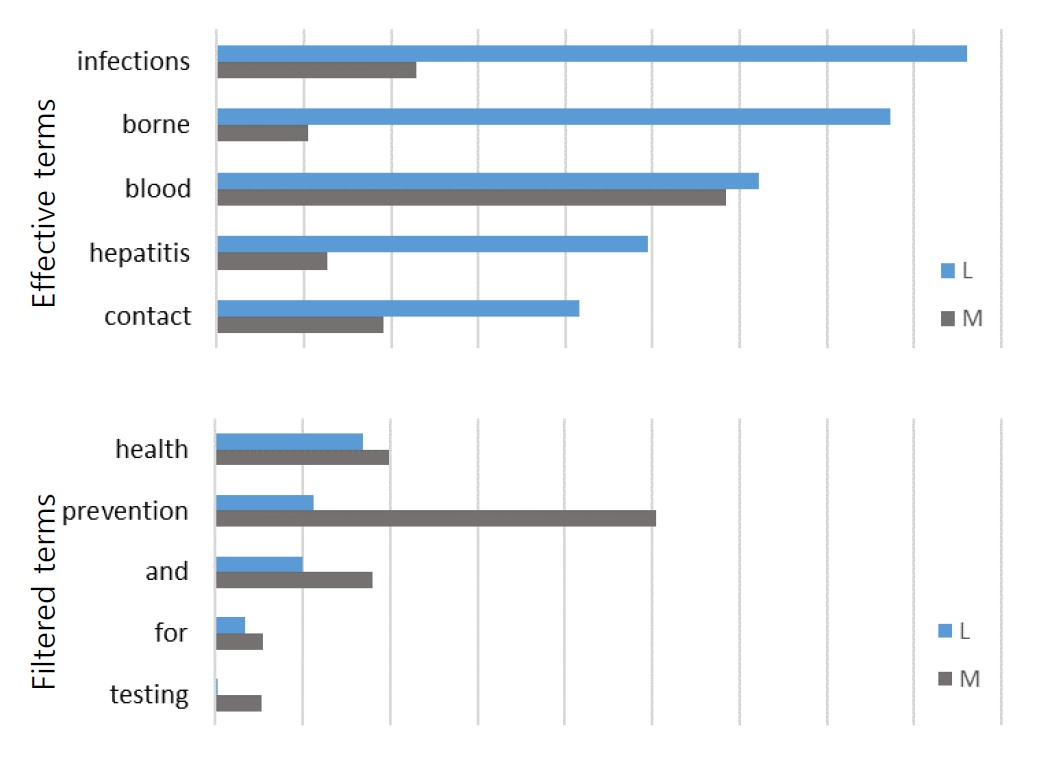}}
	\hfill
	\caption{Visualization results for the query "blood diseases that are sexually transmitted". $L$ denotes the localization map and $M$ denotes the interaction matrix. $l$ and $m$ denotes the flattened 1D-array for $L$ and $M$ correspondingly.} 
	\label{fig:heatmap}
\end{figure*}

Grad-CAM produces a localization map $L \in \mathbb{R}^{u\times v}$ which indicates the contribution of each term pair to the matching score. In order to obtain $L$, we first calculate the importance weight $\alpha_k$ of each feature map $A^k$ in the final convolutional layer. 

\begin{equation}\label{eq1}
\alpha_k = \frac{1}{Z}\sum_{i}^{}\sum_{j}^{} \frac{\partial S(Q,D)}{\partial A_{ij}^k}
\end{equation}
Importance weight $\alpha_k$ is calculated by global-average pooling the gradients over the feature map dimensions. 
$S(Q,D)$ denotes the matching score between a query $Q$ and a document $D$. $A^k$ denotes the $k$-th feature map of the last convolutional layer. $Z$ denotes the number of elements in the feature map $A^k$. 
\begin{equation}\label{eq2}
L_{Grad-CAM} = ReLU(\sum_{k}\alpha_k A^k)
\end{equation}
Localization map $L$ is calculated by applying ReLU activation function to a weighted combination of feature map activations. In equation \ref{eq2}, $L$ has the same size as $A^k$. We upsample $L$ to the input size using bilinear interpolation. Each element of the upsampled localization map, $L_{ij}$, denotes the contribution of the term pair $(q_i, d_j)$ to the matching score. $q_i$ denotes the $i$-th term in query and $d_{j}$ denotes the $j$-th term in document. Heatmap visualization of the localization map is shown in Figure \ref{fig:heatmap}. 
By calculating the cumulative sum for each column of $L$, a flattened 1D-array $l$ is obtained. Each element of the flattened array, $l_j$, denotes the cumulative contribution of document term $d_j$. By calculating the cumulative sum for each column of $M$, a flattened 1D-array $m$ is obtained. Each element of the flattened array, $m_j$, denotes the cumulative correlation between the document term $d_j$ and the given query $Q$.

\section{EXPERIMENTS}

Our experiment provides explanations for following questions. 
\begin{enumerate}
	\item Given a query and a document, which components of the document contribute to the ranking result?
	\item Given a query and a set of documents, which characteristics differentiate the ground truth document from the negative documents?
\end{enumerate}

\subsection{Experimental Setup}
We used the TREC 2019 Deep Learning Track Document Ranking Dataset\footnote{https://microsoft.github.io/TREC-2019-Deep-Learning/}. Training set is composed of 350,000 unique queries and validation set is composed of 5,000 unique queries. In both training set and validation set, each query has a positive labeled document. We used the MatchZoo\footnote{https://github.com/NTMC-Community/MatchZoo-py} implementation of MatchPyramid model and the pretrained GloVe\footnote{http://nlp.stanford.edu/data/glove.6B.zip} embedding to vectorize queries and documents.
The hyperparameters are tuned using the same setup as described in the original paper\cite{pang2016text}. The trained model achieved 0.5243 for NDCG@3, 0.593 for NDCG@5, 0.5407 for MAP.

\subsection{Evaluation Method}
To answer the questions stated above, we first conducted qualitative analysis on the localization map $L$ and the interaction matrix $M$. 
We designed an experiment to verify that $L$ can be useful to extract the most query-relevant snippet from a document. 
We performed a statistical analysis on the set of ground truth documents and the set of negative documents.

\textbf{Qualitative Analysis}
We extracted {\it effective terms} and {\it filtered terms} using the information from the flattened localization map $l$ and the flattened interaction matrix $m$. {\it Effective term} refers to the document term that contributes highly to the ranking score. {\it Filtered term} refers to the document term that do not contribute highly to the ranking score, despite its high embedding similarity with a query term. 

\textbf{Snippet Generation} 
We compared vanilla snippet generator and our Grad-CAM snippet generator. Vanilla snippet generator returns the snippet with the highest density of query terms in a fixed window size $w$. Vanilla snippet generator assigns binary value to each document term $d_j$, $1$ if the document term exactly matches a query term and 0 otherwise. 
Our Grad-CAM snippet generator adds $\frac{l_j}{w}$ to the binary value, where ${l_j}$ denotes the value of cumulative contribution of a document term ${d_j}$. We set the window size $w$ for a snippet as 20. For the 5,000 ground truth documents in validation set, we applied both types of snippet generators. From the 5,000 pairs of snippet, we sampled ten pairs that show significant difference. We conducted a survey on 87 assessors to choose a more relevant snippet given ten pairs of snippet.

\textbf{Statistical Analysis}
To answer the second question, we performed a statistical analysis on the set of ground truth documents and the set of negative documents.

\begin{enumerate}
	\item{$Kurt(L) = \mathbb{E}[({\frac{L-\mu}{\sigma}})^4]$} 
	\item{$\sum_{i}\sum_{j}L_{ij}$} 
\end{enumerate}

Given a query, we measured $Kurt(L)$ and $\sum_{i}\sum_{j}L_{ij}$ for each document. $Kurt(L)$ denotes the kurtosis, which is the fourth standardized moment.
$\sum_{i}\sum_{j}L_{ij}$ refers to the sum of contributions of every query-document term pairs given a query and a document. 
We tested the following hypothesis: Ground truth document has larger $Kurt(L)$ and $\sum_{i}\sum_{j}L_{ij}$ than the negative sample documents.

\section{RESULTS}
\subsection{Qualitative Analysis}
By investigating localization map $L$, interaction matrix $M$, flattened localization map $l$, flattened interaction matrix $m$, we were able to interpret the NRM ranking results. Figure \ref{fig:heatmap} shows the visualization of $L$, $l$, $M$, $m$, {\it effective terms}, {\it filtered terms} for the query "blood diseases that are sexually transmitted" and its ground truth document. Top image in Figure \ref{fig:heatmap}(a) shows the heat map visualization of localization map $L$, which we used "blue-red" color schema as the min-max mapping of the values. Query term "$diseases$" and document term "$infection$" was the term pair of the largest contribution to the ranking score. Bottom image in Figure \ref{fig:heatmap}(a) shows the flattened 1D-array $l$  for the localization map $L$. By calculating the vertical sum for each column in $L$, we were able to measure the cumulative contribution of each document term. Top image in Figure \ref{fig:heatmap}(b) shows the visualization of interaction matrix $M$. The darker the region, the larger the embedding similarity between term pair $(q_i, d_j)$. Query term "$diseases$" and document term "$blood$" was the term pair of the largest embedding similarity.

By comparing $L$ and $M$, we observed that the highlighted regions of each image were roughly aligned. The highlighted regions in localization map represent the {\it effective terms}. However, we detected that some regions highlighted on M were not highlighted on the corresponding location in L. Based on such observation, we defined {\it filtered terms} as the document terms that do not contribute highly to the ranking score, despite its high embedding similarity with a query term. {\it Effective terms} and {\it filtered terms} of the sample query's ground truth document are listed in Figure \ref{fig:heatmap}(c). For a document term $d_j$, the blue bar represents the value $l_j$ and the gray bar represents the value $m_j$. Document term "$infection$" is an {\it effective term} which is intuitive considering its semantic similarity with the query term "$transmitted$".
Document term "$prevention$" shows the largest difference between two values, which indicates that the NRM captures the subtle query intent of focusing on "$infection$" rather than "$prevention$".

Table \ref{tbl:effect_term} shows the {\it effective terms} and {\it filtered terms} for sampled queries. By observing {\it effective terms} and {\it filtered terms}, we could qualitatively assess the NRM's 
capability of capturing user's query intent. For instance, given the query "what kind of oil is good for dry hair", we observe that the NRM captures the user intent which focuses on dry hair rather than fragrance or aromatherapy.

\begin{table}[t]
	\caption{Examples of Effective terms and Filtered terms}
	\begin{center}
		\label{tbl:effect_term}
		\begin{tabular}{cl}
			\hline
			\multicolumn{2}{l}{Query 1: "blood diseases that are sexually transmitted"} \\
			\hline
			Effective terms &: infection, borne, contact \\ 
			Filtered terms &: prevention, health \\
			\hline 
			\multicolumn{2}{l}{Query 2: "what kind of oil is good for dry hair"} \\
			\hline
			Effective terms &: sandalwood, helps, recommend, scalp \\ 
			Filtered terms &: fragrance, aromatherapy \\
			\hline
			\multicolumn{2}{l}{Query 3: "waht language does trinidad speak"} \\
			\hline
			Effective terms &:  trinidadian, creole, language, english \\ 
			Filtered terms &: religions, dasheen  \\ 
			\hline
			\multicolumn{2}{l}{Query 4: "can hives be a sign of pregnancy"} \\
			\hline
			Effective terms &: itching, hormones, change \\ 
			Filtered terms &: insect \\
		\end{tabular}
	\end{center}  
\end{table}

\subsection{Snippet Generation}
\begin{figure}[h]
	\centering
	\includegraphics[scale=0.25]{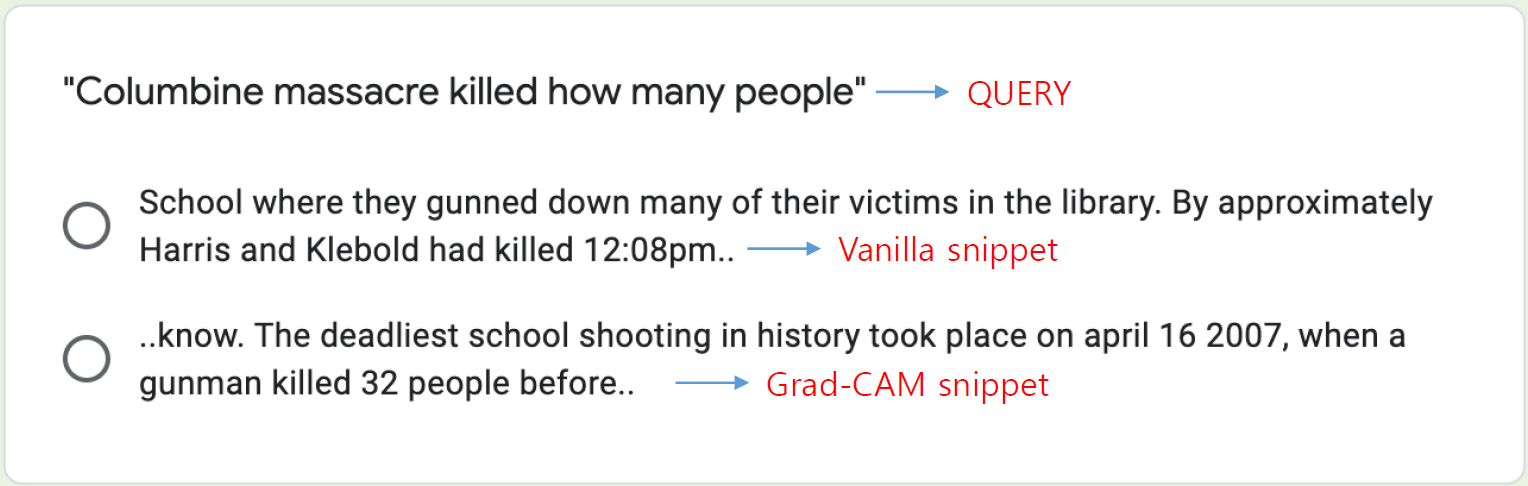}
	\caption{Sample question used in survey. Given the query "Columbine massacre killed how many people", vanilla generator produced the first snippet and our Grad-CAM generator produced the second snippet }
	\label{fig:snippet}
\end{figure}

For each of the 5,000 ground truth documents in validation set, we produced a pair of snippets using two types of snippet generation algorithm. Both vanilla generator and Grad-CAM generator produced exactly the same snippets for 83\%(4163) of the documents. For the 17\%(837) of the documents which resulted in different snippets, we sampled 10 pairs to conduct survey. Figure \ref{fig:snippet} shows a sample question of our survey, which is composed of a single query and two randomly ordered snippets.

Survey result shows that 78.37\% of the assessors chose our snippet produced from Grad-CAM generator to be more relevant. Figure \ref{fig:snippet} shows that our Grad-CAM snippet includes the main information, number of people, while the vanilla snippet highlights less relevant part. The result implies that Grad-CAM algorithm can be useful for extracting the most query-relevant snippet from the document.

\subsection{Statistical Analysis}
\begin{figure}[t]
	\centering
	\subfigure [Box plot of Kurt(L)]{\includegraphics[scale=0.22]{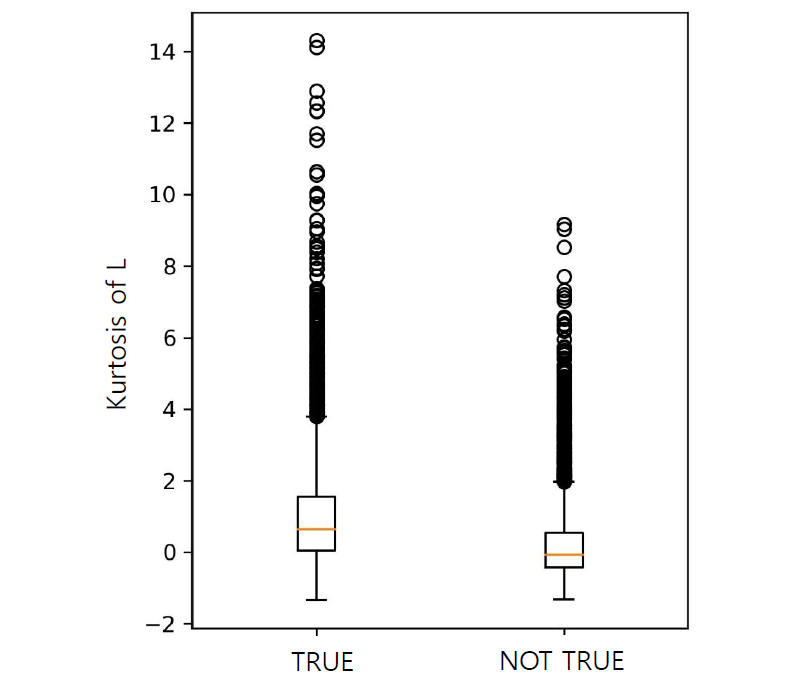}}
	\subfigure [Box plot of $\sum_{}\sum_{}L_{ij}$]{\includegraphics[scale=0.22]{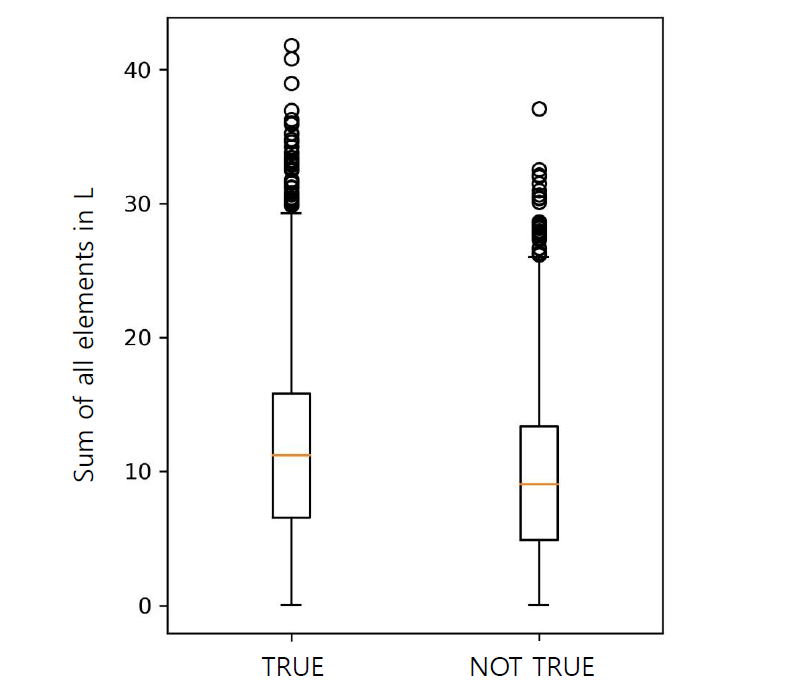}}
	\hfill
	\caption{Box plot of Kurt(L) and $\sum_{}\sum_{}L_{ij}$ for ground truth documents and negative documents. Kurt(L) denotes the kurtosis of $L$. $\sum_{}\sum_{}L_{ij}$ denotes the sum of elements in $L$.}
	\label{fig:kurtosis}
\end{figure}

By comparing the visualization results for ground truth document and negative documents, we observed that the distributions showed difference in size of tail and area under curve. This led us to perform statistical analysis on two measures: $Kurt(L)$ and $\sum_{i}\sum_{j}L_{ij}$. For the 5,000 ground truth documents in validation set, we calculated the values for each measure.  Figure \ref{fig:kurtosis} shows the box plot of two measures. We performed Mann–Whitney U test to test the following hypothesis: Ground truth document has larger $Kurt(L)$ and $\sum_{i}\sum_{j}L_{ij}$ than the negative sample documents. The p-value, smaller than 0.0001, indicates that the hypothesis holds true for both measures.

\section{DISCUSSION and future work}
In this paper, we propose an NRM interpretation method by adopting Grad-CAM algorithm. Grad-CAM algorithm is a gradient-based attribution method which has been verified through sanity checks. Our method provides explanations through qualitative analysis, snippet generation, and statistical analysis. Our method can quantify the contributions of each query-document term pair and the cumulative contribution of each document term. 
We extracted {\it effective} {\it terms} and {\it filtered} {\it terms} for each document. Snippet generation is a practical application of our method to extract the most query-relevant snippet from a document. 
Statistical analysis proves that the ground truth documents have larger values for kurtosis and sum of elements in $L$, when compared to the negative sample documents. 
In the future work, we would like to utilize these measures as neural ranking model features. 
Furthermore, since Grad-CAM algorithm is limited to CNN architectures, we would like to generalize our method to other gradient-based methods.



\bibliographystyle{ACM-Reference-Format}
\bibliography{acmart}


\end{document}